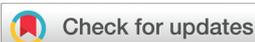

Check for updates

# Nonlinear ferroelectric characteristics of barium titanate nanocrystals determined *via* a polymer nanocomposite approach†

Qiong Li,[a] Elshad Allahyarov, [ID] [b,c,d] Tianxiong Ju, [ID] [a] Zhiqun Lin [ID] *[e] and Lei Zhu [ID] *[a]

The growing demand for high energy storage materials has garnered substantial attention towards lead-free ferroelectric nanocrystals (NCs), such as BaTiO₃ (BTO), for next-generation multilayer ceramic capacitors. Notably, it remains challenging to accurately measure the dielectric constant and polarization−electric field (*P–E*) hysteresis loop for BTO NCs. Herein, we report on nonlinear ferroelectric characteristics of BTO NCs *via* a polymer nanocomposite approach. Specifically, poly(vinyl pyrrolidone) (PVP)/BTO nanocomposite films of 3–10 μm thickness, containing 380 nm tetragonal-phased and 60 nm cubic-phased BTO NCs with uniform particle dispersion, were prepared. Theoretical deconvolution of the broad experimental *P–E* loops of the PVP/BTO NC composite films revealed three contributions, that is, the linear deformational polarization of the nanocomposites, the polarization of BTO NCs ($P_p$), and the polarization from strong particle–particle interactions. Using different mixing rules and nonlinear dielectric analysis, the overall dielectric constants of BTO NCs were obtained, from which the internal field in the BTO NCs ($E_p$) was estimated. Consequently, the $P_p$–$E_p$ hysteresis loops were obtained for the BTO380 and BTO60 NCs. Interestingly, BTO380 exhibited square-shaped ferroelectric loops, whereas BTO60 displayed slim paraelectric loops. This work presents a robust and versatile route to extract the $P_p$–$E_p$ loops of ferroelectric NCs from polymer/ceramic nanocomposites.

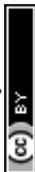

## Introduction

High dielectric constant ferroelectric nanocrystals (NCs), especially lead-free BaTiO₃ (BTO), are highly desirable for manufacturing high-capacitance-density multilayer ceramic capacitors.[1–5] However, as the size of perovskite NCs decreases below a threshold, both surface and bulk defects become dominant, preventing the formation of the tetragonal phase and ferroelectric domains.[6–8] Such a critical size was found to depend largely on different synthesis methods and post-thermal annealing processes, ranging from 85 to

*[a]Department of Macromolecular Science and Engineering, Case Western Reserve University, Cleveland, Ohio 44106, USA. E-mail: lxz121@case.edu*
*[b]Institut für Theoretische Physik II: Weiche Materie, Heinrich-Heine Universität Düsseldorf, Universitätsstrasse 1, 40225 Düsseldorf, Germany*
*[c]Theoretical Department, Joint Institute for High Temperatures, Russian Academy of Sciences (IVTAN), 13/19 Izhorskaya Street, Moscow 125412, Russia*
*[d]Department of Physics, Case Western Reserve University, Cleveland, Ohio 44106-7079, USA*
*[e]Department of Chemical and Biomolecular Engineering, National University of Singapore, Singapore 117585, Singapore. E-mail: z.lin@nus.edu.sg*
†Electronic supplementary information (ESI) available. See DOI: https://doi.org/10.1039/d3nr05185d

300 nm.[7,9–14] For sizes below the threshold value, BTO NCs usually exhibit a paraelectric cubic phase with a low dielectric constant.

Recently, we investigated the structure–dielectric property relationship for a series of combustion-made BTO NCs with the average size ranging from 50 to 500 nm.[15] Several findings were obtained from this study. First, the critical size was between 100 and 200 nm for such combustion-made BTO NCs. With a particle size of ≥200 nm, ferroelectric BTO NCs contained an ~85% tetragonal phase and an ~15% cubic fraction. With a particle size of ≤100 nm, paraelectric BTO NCs exhibited a tetragonal-fluctuated cubic phase with large microstrains, indicating a significant lattice distortion from defects. Second, the ferroelectric BTO NCs displayed a relatively high dielectric constant of 150–300, whereas the paraelectric NCs had a low dielectric constant only around 80. Nevertheless, these values are much lower than the dielectric constants (a few thousands) of bulk (*i.e.*, sintered) BTO samples. The much lower dielectric constant of BTO NCs was primarily attributed to the absence of 90° domains and domain walls in single crystalline BTO NCs. Third, this "size effect" was considered to be a result of surface and/or bulk defects trapped in the samples. However, it has been difficult to pinpoint the types and locations of these defects using high-resolution transmission electron microscopy.

In addition to the linear dielectric properties, the nonlinear ferroelectric properties of BTO NCs have not been determined directly and accurately. Although piezoresponse force microscopy (PFM) can measure piezoelectric responses and phase hysteresis loops,[16] it cannot be used directly to obtain electric displacement–electric field ($D$–$E$) loops. In particular, the parallel capacitor geometry is difficult to obtain for sphere-shaped NCs. Moreover, even though a parallel capacitor geometry has been achieved for nanosized perovskite thin films, significant stray capacitance exists in the measurement, which needs to be subtracted to produce accurate $D$–$E$ loops.[17,18]

Herein, we report a nanocomposite approach to determine the $D$–$E$ loops of BTO NCs. Using a theoretical deconvolution of the experimental $D$–$E$ loops based on the Langevin-type function,[19] three contributions were identified for poly(vinyl pyrrolidone) (PVP)/BTO NC composites, that is, linear deformational polarization, BTO NC polarization, and particle–particle dipolar interaction. Specifically, two sets of PVP/BTO composites were prepared, namely, PVP/BTO380 (particle size of ~380 nm with the ferroelectric tetragonal phase) and PVP/BTO60 (particle size of ~60 nm with the paraelectric cubic phase). The ferroelectric properties of the BTO380 and BTO60 NCs were thoroughly compared. Notably, BTO380 and BTO60 were found to display a square-shaped ferroelectric loop and a slim paraelectric loop, respectively. Both nanocomposites manifest a strong particle–particle interaction at a high packing fraction, accounting for the significant nonlinear dielectric loss.

# Experimental section

## Materials

BTO NCs (99.9% purity) were purchased from U.S. Research Nanomaterials Inc. (Houston, TX). Poly(vinyl pyrrolidone) (average molecular weight of 58 kDa, Acros Organics) and ethanol were purchased from Fisher Scientific (Waltham, MA) and used without purification. Multi-domain BaTiO$_3$ single crystals (BTO SCs) with the (001) orientation (5 mm × 5 mm × 0.5 mm, both sides polished) were purchased from MSE Supplies, LLC (Tucson, AZ).

The paraelectric BTO NCs had an average particle size of 62 ± 16 nm (mean ± standard deviation),[15] and are denoted as BTO60. The as-received 200 nm (218 ± 65 nm) ferroelectric BTO NC powder was placed in a 10 mL alumina crucible and heated in a high-temperature box furnace (SentroTech, Strongsville, OH) at 950 °C for up to 72 h to improve its ferroelectric properties by removing the structural defects. After thermal annealing, the particle size increased to 377 ± 102 nm due to the Ostwald ripening. This ferroelectric BTO NC sample is denoted as BTO380. The particle sizes were determined by field-emission scanning electron microscopy (SEM) over 500 particles by manually measuring each particle size using ImageJ software.

## PVP/BTO nanocomposite film preparation *via* spin-coating

PVP films with different volume fractions ($\eta$) of BTO60 and BTO380 NCs, *i.e.*, around 20, 30, and 40 vol%, were prepared by spin-coating. First, PVP was dissolved in 2.0 mL of ethanol to prepare a 10 wt% polymer solution. The BTO NCs were added to PVP solution to prepare suspensions with different BTO volume fractions. The suspension was stirred for at least 1 h at room temperature and ultrasonicated for about 20 min in a sonication bath (300 W) prior to spin-coating. As reported earlier,[19] relatively uniform suspensions were obtained with PVP/ethanol *via* ultrasonication.

Clean microscope slides (2.5 cm × 2.5 cm) were sputter-coated with a thin layer (10 nm) of gold (Au) on both sides using a Quorum Q300T D Plus sputter coater (Quorum Technologies, Laughton, East Sussex, U.K.). The dispersed PVP/BTO suspension was then spin-coated on the Au-coated microscope slides at 500 revolutions per minute (rpm) for 20 s and then at 1000 rpm for 40 s at an acceleration rate of 500 rpm s$^{-1}$. After spin-coating, the PVP/BTO nanocomposite films were dried in a vacuum at 80 °C overnight. Then, circular Au electrodes with an area of 5.15 mm$^2$ were sputter-coated on the film top surface. The final film thicknesses were measured using a P-6 stylus profilometer (KLA-Tencor Corporation, Milpitas, CA). These films were stored in a vacuum desiccator before use.

## Characterization

Thermogravimetric analysis (TGA) was performed using a TA Instruments Q500 at a heating rate of 10 °C min$^{-1}$ up to 700 °C under a dry N$_2$ atmosphere. Approximately, 5 mg of the film sample was placed in a platinum pan for each measurement. Based on the mass loss and densities of BTO (6.02 g cm$^{-3}$) and PVP (1.20 g cm$^{-3}$), the volumetric packing fractions of BTO NCs in the PVP/BTO380 nanocomposites were 0.23, 0.34, and 0.40, while those in the PVP/BTO60 nanocomposites were 0.21, 0.31 and 0.41, respectively. Therefore, these film samples are denoted as PVP/BTO380-0.23, PVP/BTO380-0.34, PVP/BTO380-0.40, PVP/BTO60-0.21, PVP/BTO60-0.31, and PVP/BTO60-0.41, respectively.

A Thermo Fisher Apreo 2 field-emission SEM, operating at 10 kV and 0.1 nA, was used to examine the particle sizes and distribution of BTO NCs in the PVP/BTO nanocomposite films. The operation mode was immersion and a Trinity T1 detector was used to collect the back-scattered electrons and minimize charging from secondary electrons.

Conventional two-dimensional (2D) X-ray diffraction (XRD) patterns were obtained using a Rigaku MacroMax 002$^+$ instrument equipped with a Confocal Max-Flux optic and a microfocus X-ray tube source, operating at 45 kV and 0.88 mA. The Cu Kα X-ray wavelength was 1.5418 Å. 2D WAXD patterns were recorded using a Fujifilm image plate and scanned with an Amersham™ Typhoon scanner (Cytiva, Marlborough, MA) at a resolution of 100 μm per pixel. The typical acquisition time was 12 h. One-dimensional (1D) WAXD curves were obtained by integrating the corresponding 2D WAXD patterns radially using the Polar software developed by Stonybrook Technology and Applied Research, Inc. (Stony Brook, NY).

Polarization–electric field ($P$–$E$) loop measurements were performed on a Premiere II ferroelectric tester (Radiant Technologies, Inc., Albuquerque, NM) equipped with a Trek 10/10B-HS high voltage amplifier (0–10 kV AC, Lockport, NY). For the BTO (001)-SC sample, circular Au electrodes with an area of 7.06 mm$^2$ were sputter-coated on both sides of the disk sample. The Au-coated film of the PVP/BTO nanocomposites was immersed in a silicone oil bath and annealed at 110 °C overnight. Prior to the $P$–$E$ loop tests, it was cooled to room temperature to avoid moisture uptake. During the $P$–$E$ loop test, the samples were immersed in silicone oil to avoid corona discharge in air.

## Results and discussion

### Fabrication of PVP/BTO nanocomposite films with uniform particle dispersion

The crystalline structures of the BTO380 and BTO60 NCs were determined by XRD, as shown in Fig. 1A and E. From peak assignments, the BTO380 NCs exhibited a ferroelectric tetragonal phase and the BTO60 NCs exhibited a paraelectric cubic phase. As reported earlier,[19] the BTO NCs could be dispersed with PVP in ethanol via ultrasonication. The Fourier transform infrared result suggested strong dipole–ion interactions between the amide groups in PVP and the Ba$^{2+}$/Ti$^{4+}$ ionic species on the BTO surfaces. As a result, PVP formed an adsorbed corona layer around the BTO core, stabilizing the BTO NCs in ethanol through the steric repulsion effect. Therefore, in this study PVP was chosen among other polymers as the polymer matrix. PVP/BTO nanocomposite films with uniform particle distribution were fabricated by spin-coating, and the average thicknesses of final films were determined using a stylus profilometer: 4.66 ± 0.16 μm for PVP/BTO380-0.23, 3.87 ± 0.20 μm for PVP/BTO380-0.34, 6.60 ± 0.20 μm PVP/BTO380-0.40, 3.20 ± 0.08 μm for PVP/BTO60-0.21, 3.55 ± 0.12 μm for PVP/BTO60-0.31, and 6.70 ± 0.52 μm for PVP/BTO60-0.41, respectively. The film thickness should be at least 3 μm to avoid the formation of pinholes penetrating through the sample. Also, thick films (>10 μm) could not be easily obtained by spin-coating. SEM micrographs of the PVP/BTO380 and PVP/BTO60 nanocomposite films are shown in Fig. 1B–D and F–H, respectively. The bright spots in the micrographs were the BTO NCs embedded in the dark PVP matrix. The average particle sizes of BTO NCs were determined by counting over 500 particles for BTO dry powders in the SEM images: 377 ± 102 nm for BTO380 and 62 ± 16 nm for BTO60.[15] By employing the immersion mode with a T1 detector for back-scattered electrons, the BTO NCs were readily visible due to their large atomic number contrast to that of the PVP matrix. In this way, charging from secondary electrons was largely avoided. No large BTO particle aggregates were observed, indicating the uniform dispersion of BTO NCs in the PVP matrix, even though the η of ca. 0.40 exceeded the percolation threshold for spherical particles (~35 vol%[20,21]). The uniform dispersion of BTO NCs in the PVP matrix was important for polymer nanodielectrics to avoid easy dielectric breakdown.[6] There were no macroscopic cracks found in the Au-coated electrode areas, which would cause high leakage and early breakdown during the P–E loop test. The BTO380 NCs exhibited primarily the tetragonal phase (~87.5% as determined previously,[15] see Fig. 1A) because the diffractogram clearly showed peak splitting for the (002)/(200) reflections. Meanwhile, the (200) reflection of BTO60 appeared to be broad and symmetric, suggesting that BTO60 was in the paraelectric cubic phase with a large microstrain (see Fig. 1E).[15] From the TGA thermograms (Fig. S1 in the ESI†), the actual η of BTO NCs was determined by calculating the weight loss of PVP in the PVP/BTO composites.

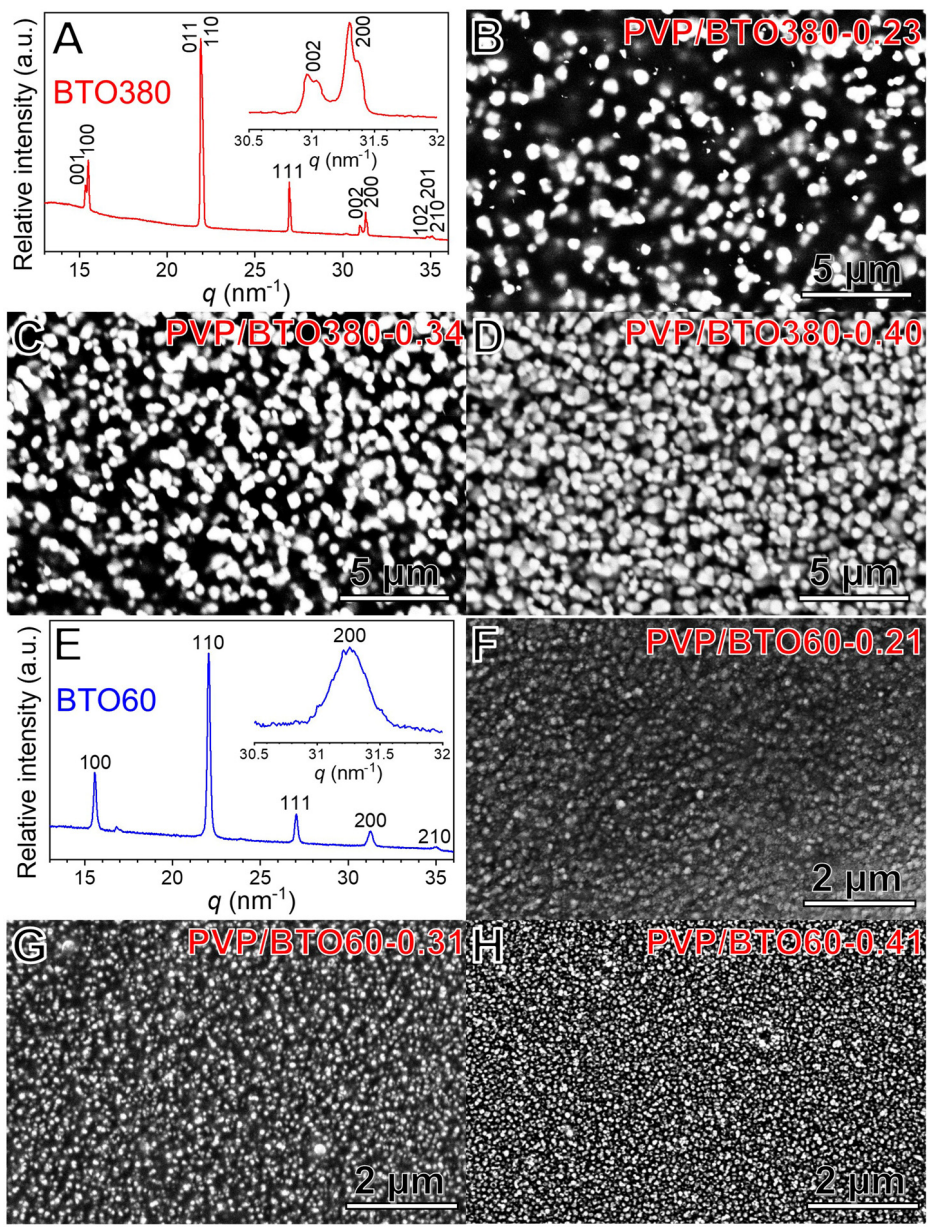

**Fig. 1** Powder XRD patterns of (A) BTO380 and (E) BTO60 with the insets showing the enlarged 30.5−32 nm$^{-1}$ region. SEM micrographs of (B) PVP/BTO380-0.23, (C) PVP/BTO380-0.34, (D) PVP/BTO380-0.40, (F) PVP/BTO60-0.21, (G) PVP/BTO60-0.31, and (H) PVP/BTO60-0.41.

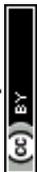

## Deconvolution of experimental P–E loops for PVP/BTO380 nanocomposites

A series of continuous bipolar poling cycles were run at 10 Hz at room temperature under different electric fields with a sinusoidal waveform. At each poling field, five continuous bipolar loops were recorded, and the second P–E hysteresis loop is presented in Fig. 2A, C and E for the PVP/BTO380 composites with η = 0.23, 0.34, and 0.40, respectively. The second and the fifth loops largely overlapped, indicating that the steady state was achieved after the second loop. The AC electronic conduction under high electric fields and a small permanent remanent polarization ($P_{r0}$) were subtracted following our previous reports.[22,23] Ferroelectric-shaped hysteresis loops were observed at high poling electric fields, especially for the composites with high $\eta$ values. It is natural to consider that this ferroelectric behavior originates from the ferroelectricity of the tetragonal BTO380 NCs. When the $\eta$ was below 0.34, the PVP/BTO380 composite films could be poled up to 80 MV m$^{-1}$ at room temperature. However, the PVP/BTO380-0.40 film could only survive at a lower poling field of 46 MV m$^{-1}$ because of a significant reduction in the breakdown strength at such a high $\eta$ value. Under a similar electric field, namely 52 MV m$^{-1}$ for PVP/BTO380-0.23, 49 MV m$^{-1}$ for PVP/BTO380-0.34, and 46 MV m$^{-1}$ for PVP/BTO380-0.40, the spontaneous polarization, $P_{s,film}$, increased from 3.9 mC m$^{-2}$ to 6.8 and 32.8 mC m$^{-2}$, respectively. Additionally, the coercive field $E_c$ slightly increased from 15.9 MV m$^{-1}$ to 17.6 and 18.1 MV m$^{-1}$. When the $\eta$ value was as high as 0.40, square-shaped hysteresis loops were observed.

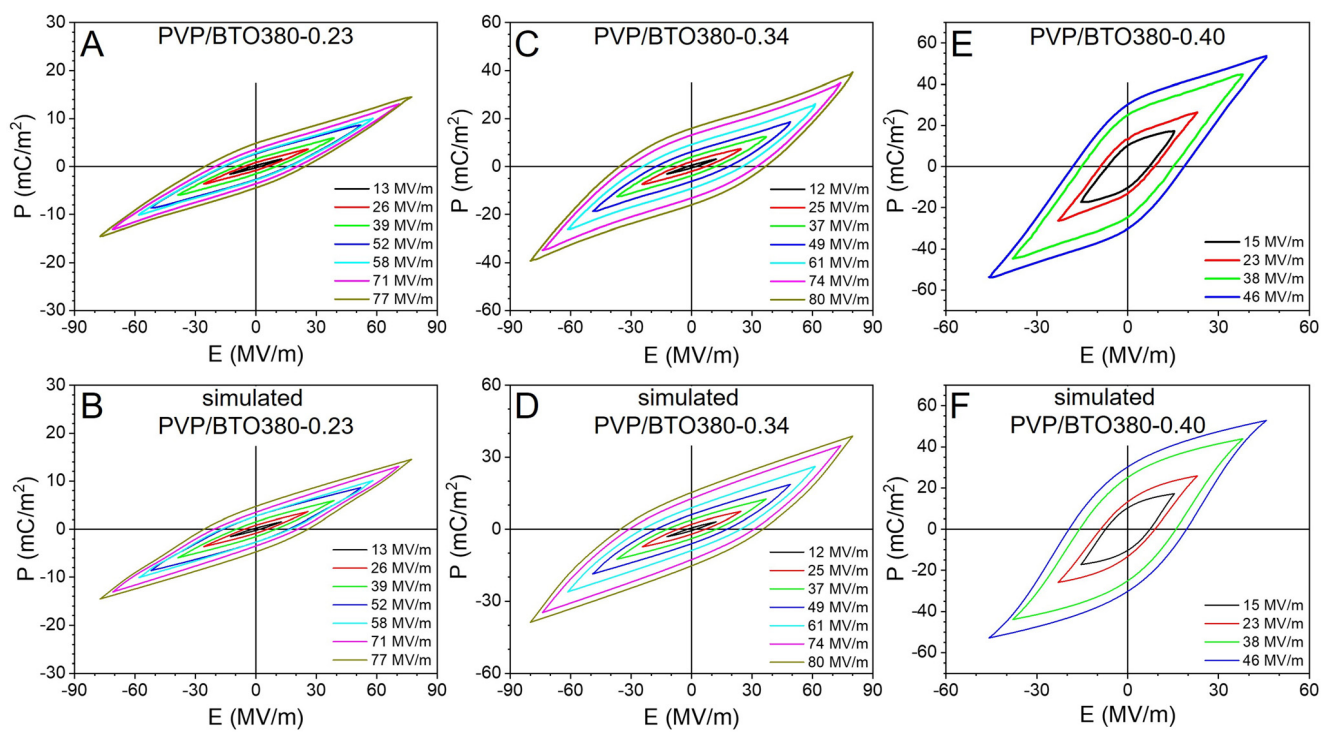

**Fig. 2** Bipolar P–E loops of the PVP/BTO380 composites with different BTO packing fractions at room temperature: (A) PVP/BTO380-0.23, (C) PVP/BTO380-0.34, and (E) PVP/BTO380-0.40. The corresponding simulated P–E loops are shown in (B), (D), and (F), respectively. The poling frequency is 10 Hz with a sinusoidal waveform.

Based on the Langevin-type functions,[19] the broad experimental $P$–$E$ loops of the polymer/BTO composites could be deconvoluted into three components: (i) the linear and deformational polarization of the entire composite $P_L$, (ii) BTO particle polarization $P_p(E)$, and (iii) particle–particle dipolar interaction $P_{int}(E)$. In this work, the $P$–$E$ loop deconvolution was performed and the fitted overall $P$–$E$ loops for the PVP/BTO380 composites are shown in Fig. 2B, D and F. The simulated loops fitted well with the experimental results.

The deconvoluted linear $P_L$–$E$ loops of various PVP-BTO380 composites are presented in Fig. 3A–C. The linear dielectric constants ($\varepsilon_{c1,L}$) of the composites were obtained from the slope of the $P_L$–$E$ loops and are summarized in Fig. 3D. For each composite, the $\varepsilon_{c1,L}$ slightly increased with increasing poling electric field. This is because the deformational polarization is largely determined from the $\eta$ of the composites, not from the external poling field. Indeed, as we can see from Fig. 3D, $\varepsilon_{c1,L}$ increased substantially with increasing $\eta$ under the same electric field. This can be explained by the mixing rules of composite materials, which will be discussed later.

Two nonlinear contributions, $P_p(E)$ and $P_{int}(E)$, were extracted from the nonlinear component of the $P$–$E$ loops of the PVP/BTO380 composites and are shown in Fig. 4 and 5, respectively. To determine which loop was which, we need to treat three loops of the PVP/BTO380 composites with different filler contents together under the same or similar poling electric field following our previous report.[19] Specifically, the $P_p$–$E$ loop should remain constant for the same ferroelectric particles in different composites. The $P_{int}$–$E$ loops should gradually increase its maximum polarization ($P_{max}$) and broaden as the filler content increased. Nonlinear dielectric analysis was carried out to obtain apparent nonlinear dielectric constants following previous reports.[24,25] The procedure is shown in

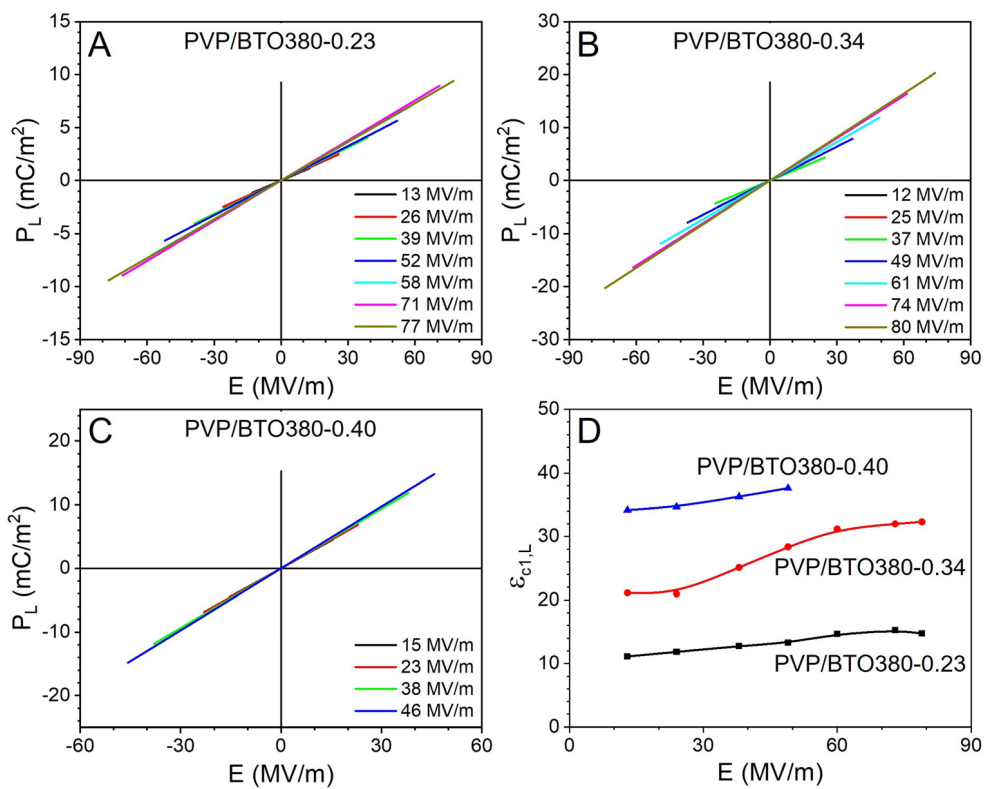

**Fig. 3** Deconvoluted linear $P_L$–$E$ loops of various PVP/BTO380 composites: (A) PVP/BTO380-0.23, (B) PVP/BTO380-0.34, and (C) PVP/BTO380-0.40 under different poling electric fields. (D) Linear dielectric constants ($\varepsilon_{c1,L}$) of various PVP/BTO380 composites calculated from the corresponding linear $P_L$–$E$ loops.

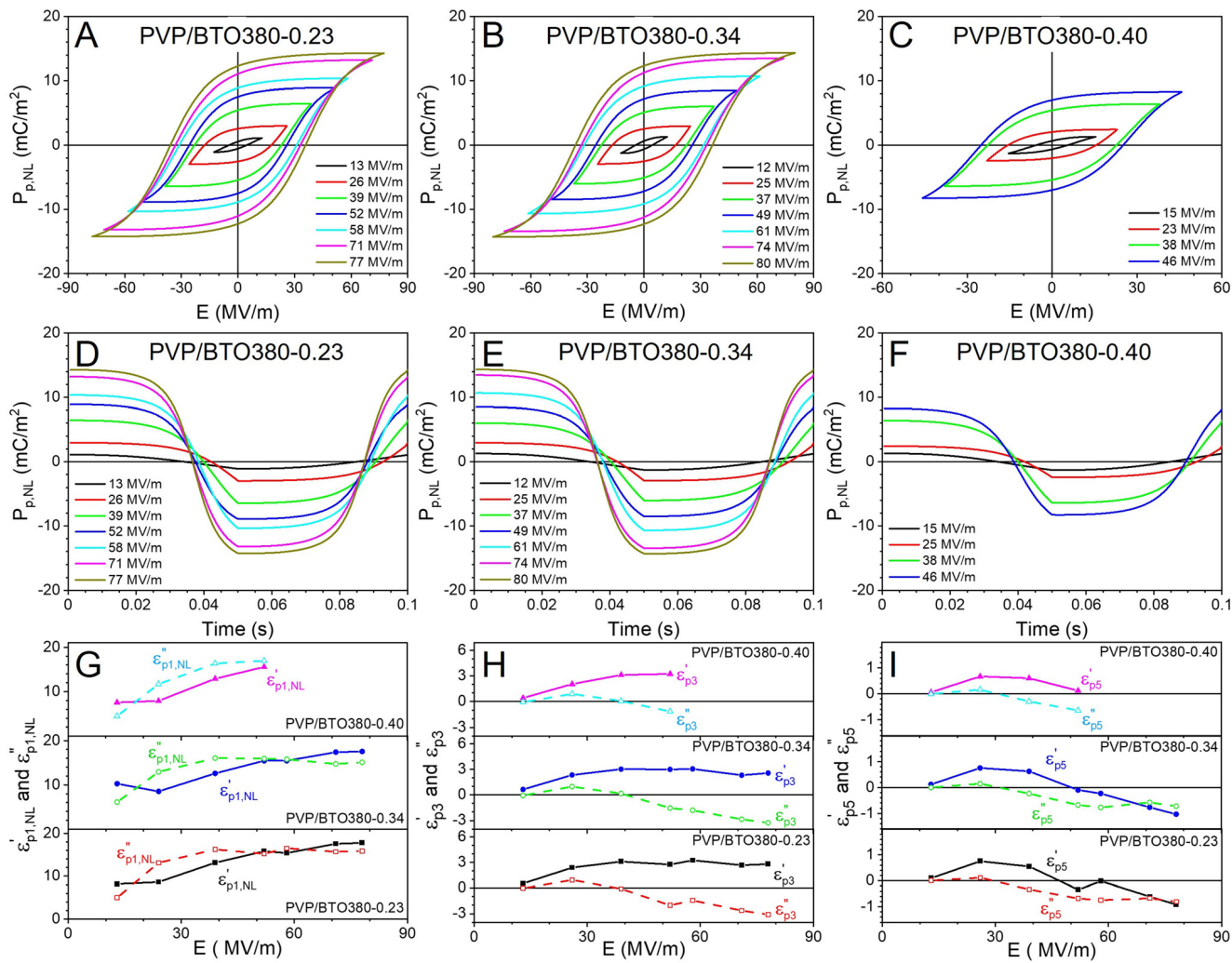

**Fig. 4** (A–C) Bipolar $P_p$–$E$ loops of BTO380 NCs extracted from the $P$–$E$ loops of the PVP/BTO380 composites with different BTO packing fractions. (D–F) Nonlinear $P_p(t)$ waves as a function of time for neat BTO380 NCs. After Fourier transform, the nonlinear harmonics, (G) $\varepsilon'_{p1,NL}$ and $\varepsilon''_{p1,NL}$ (H) $\varepsilon'_{p3}$ and $\varepsilon''_{p3}$, and (I) $\varepsilon'_{p5}$ and $\varepsilon''_{p5}$, were calculated from the nonlinear $P_p(t)$ waves.

Scheme S1.† Briefly, an experimental $D$–$E$ loop of the PVP/BTO380 composite was first deconvoluted into a linear (or deformational) component $D_{def}$ (see Fig. 3 and the corresponding discussion) and a nonlinear component $P_{NL}$. Then, $P_{NL\pm}(E)$ was converted to a function of time, $P_{NL\pm}(t)$. After the Fourier transformation of $P_{NL}(t)$, the total $D_n^*$, linear $D_L^*$, and nonlinear $D_n^{NL*}$ ($n = 1, 2, 3, 4$ and 5) polarizations were obtained. For a dielectric with $P_{r0} = D_0 = 0$, the even-numbered harmonics are zero. Finally, the nonlinear dielectric constant $\varepsilon_{pn}^{NL*}$ can be calculated from equation $\varepsilon_{pn}^{NL*} = D_n^{NL*}/\varepsilon_0 E_0$ ($\varepsilon_0$ is the vacuum permittivity).

The deconvoluted $P_p(E)$s for the BTO380 NCs from the $P$–$E$ loops of various PVP/BTO380 composites are shown in Fig. 4A–C. Typical ferroelectric loops were observed. Note that the electric field in the $x$-axis was the applied electric field on the composite, not the local field inside the BTO380 NCs. From our recent study,[15] the BTO380 NCs were largely single crystals after thermal annealing at 950 °C. After high-field elec-tric poling, single-domain particles were likely to be obtained. Therefore, these $P_p(E)$ loops represented the ferroelectric hysteresis loops for randomly oriented single-domain, single-crystal BTO nanoparticles. Because $P_p(E)$ was for individual BTO380 NCs and it did not consider the neighboring particles, this contribution did not depend on the $\eta$ of the nanocomposites. This is exactly seen in Fig. 4A–C. For example, the $P_{p,NL}$–$E$ loops are around 50 MV m$^{-1}$ for various PVP/BTO380 composites, the maximum polarization $P_{max}$ is $\sim$ 8.6 mC m$^{-2}$, the remanent polarization $P_r$ is $\sim$ 7.2 mC m$^{-2}$, and the coercive field $E_c$ is $\sim$ 26 MV m$^{-1}$.

The corresponding $P_p(t)$ curves are shown in Fig. 4D–F. After Fourier transformation, nonlinear polarizations, $D_n^{NL*}$, were calculated. When $n > 5$, the values of nonlinear dielectric constants became very small and were neglected. Even-numbered ($n = 2, 4$) nonlinear dielectric constants were zero because the samples did not exhibit any $P_{r0}$.[24–26] The odd-numbered ($n = 1, 3, 5$) nonlinear dielectric constants $\varepsilon_{pn}^{NL*}$ of various PVP/BTO380 composites are shown in Fig. 4G–I. As the most significant contributors to nonlinearity, both $\varepsilon'_{p1,NL}$ and $\varepsilon''_{p1,NL}$ increased with increasing electric field for all compositions. For $n = 3$, the $\varepsilon'_{p3}$ values were positive and continuously increased with increasing electric field, while $\varepsilon''_{p3}$ first increased at lower fields and then decreased to below zero for all compositions when $E > 39$ MV m$^{-1}$. $\varepsilon''_{p5}$ showed the same trend as $\varepsilon''_{p3}$, but $\varepsilon'_{p5}$ initially increased when $E < 39$ MV m$^{-1}$ and then started to decrease at fields above 39 MV m$^{-1}$ for all compositions. Please note that all these nonlinear harmonics contribute to the dielectric loss.

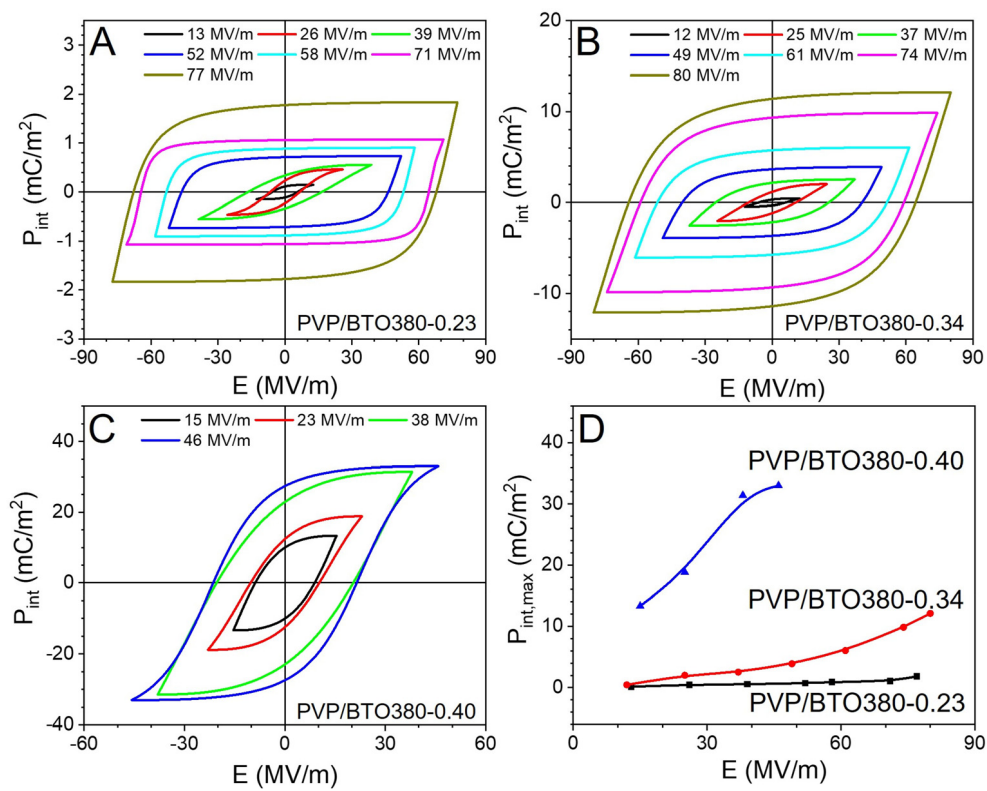

**Fig. 5** (A–C) Bipolar particle–particle interaction $P_{int}$–E loops extracted from the experimental P–E loops of the PVP/BTO380 composites with different BTO packing fractions. (D) The corresponding maximum polarization ($P_{int,max}$) under different electric fields.

The particle–particle interaction $P_{int}(E)$ loops of various PVP/BTO380 composites with $\eta = 0.23$, 0.34 and 0.40 are presented in Fig. 5A–C. Typical ferroelectric-shaped loops were observed. The maximum $P_{int}$ ($P_{int,max}$) values are plotted in Fig. 5D. At low packing fractions (e.g., $\eta = 0.23$), $P_{int,max}$ remained low and did not increase much with increasing electric field. This is because of the large average interparticle distance of 362 nm (assuming a face-centered cubic packing). For $\eta = 0.34$ and the average interparticle distance decreased to 225 nm, $P_{int,max}$ gradually increased with increasing electric field. For $\eta = 0.40$ and the average interparticle distance became 173 nm, $P_{int,max}$ increased substantially with increasing electric field. It is also likely that as the $\eta$ increased, the BTO380 NCs tended to flocculate locally, further increase the particle–particle dipolar interactions. Different from the extracted BTO380 particle loops, which were independent of the $\eta$, $P_{int,max}$ had obvious strong dependence on $\eta$. For example, $P_{int,max}$ increased from 0.73 mC m$^{-2}$ to 3.89 mC m$^{-2}$ and 33.02 mC m$^{-2}$ at 50 MV m$^{-1}$ with $\eta = 0.23$, 0.34 and 0.40, respectively. $P_{int}$ contribution was roughly the same as $P_p$ at $\eta = 0.34$ and significantly higher than $P_p$ in a high $\eta$ of 0.40. This result suggested that the particle–particle dipolar interaction contributed much more to the broadness of the nanocomposite hysteresis loops than the BTO particle ferroelectric switching. These nonlinear $P_{int}$–E hysteresis loops resulted in a significant dielectric loss, which is undesired for capacitive energy storage applications. Therefore, the particle–particle dipolar interaction should be mitigated for the nanodielectrics, especially with a large permittivity contrast and at a high filler content.

The permittivity of BTO380 NCs should contain two contributions, i.e., linear and nonlinear components. The linear dielectric constant $\varepsilon_{p,L}$ can be obtained from the linear dielectric constant of the PVP/BTO380 composites $\varepsilon_{c,L}$ using different mixing rules. In this work, three mixing rules were used to calculate the $\varepsilon_{p,L}$ of BTO380 NCs. The Bruggeman effective approximation is given by:[27,28]

$$\eta \frac{\varepsilon_{p,L} - \varepsilon_{c,L}}{\varepsilon_{p,L} + 2\varepsilon_{c,L}} + (1 - \eta) \frac{\varepsilon_{PVP} - \varepsilon_{c,L}}{\varepsilon_{PVP} + 2\varepsilon_{c,L}} = 0 \qquad (1)$$

where $\varepsilon_{c,L}$, $\varepsilon_{p,L}$, and $\varepsilon_{PVP}$ ($\varepsilon_{PVP}$ = 6.8 (ref. 19)) are linear dielectric constants of the PVP/BTO composites, the BTO380 NCs, and the PVP matrix, respectively. In our previous publications, we discussed that the Bruggeman mixing rule could fit with the experimental data (which usually contain particle–particle interactions) better than the Maxwell–Garnett mixing rule.[6,29] The Looyenga and Birchak mixing rules are given by:

$$\varepsilon_{c,L}{}^{\gamma} = \eta\varepsilon_{p,L}{}^{\gamma} + (1-\eta)\varepsilon_{PVP}{}^{\gamma} \tag{2}$$

where $\gamma = 1/3$ for the Looyenga equation and $\gamma = 1/2$ for the Birchak equation.[30] These mixing rules do not use any physical models and are only mathematical fitting equations. They were used to compare with the Bruggeman mixing rule. Note that if the Maxwell–Garnett mixing rule was used, the $\varepsilon_{p,L}$ values were calculated to be negative; therefore, it could not be used in this study.

Using these mixing rules, the $\varepsilon_{p,L}$ values of BTO380 NCs under various applied electric fields are plotted in Fig. 6A. Note that the Bruggeman equation was not be used for the $\eta = 0.23$ composite because no solution for the permittivity of BTO380 NCs could be obtained. Among these fittings, the highest $\varepsilon_{p,L}$ value was obtained for the Bruggeman mixing rule and the lowest $\varepsilon_{p,L}$ value was obtained for the Birchak equation.[30] The $\varepsilon_{p,L}$ values obtained using the Bruggeman model ranged from 101 to 247 with a strong linear dependency when the applied electric field increased from 13 to 78 MV

m$^{-1}$. Meanwhile, the $\varepsilon_{p,L}$ values ranged from 66 to 142 for the Looyenga model and from 54 to 105 for the Birchak model. These values seemed to be lower than the commonly reported values for the ferroelectric BTO NCs, as we reported recently.[15] Therefore, the Bruggeman mixing rule should be more suitable to estimate the dielectric constants of BTO380 NCs.

Then, the nonlinear dielectric constant was obtained by the summation of $\varepsilon_{p1,NL}$, $\varepsilon_{p3}$, and $\varepsilon_{p5}$ (see Fig. 4G–I). Combining both linear and nonlinear dielectric constants, the total dielectric constants of BTO380 NCs are plotted in Fig. 6B–D with various mixing rules. Using the Bruggeman model, the $\varepsilon_p$ values ranged from 110 to 275 when the applied electric field increased up to 77 MV m$^{-1}$, regardless of different $\eta$ values. This is because the linear $\varepsilon_{p,L}$ was the major contributor to the total permittivity of BTO380 NCs and it was more or less independent on the $\eta$. Although the nonlinear $\varepsilon_{p,NL}$ was dependent on the $\eta$, it was a minor contributor and would not affect the overall $\varepsilon_p$ much.

To obtain the $P$–$E$ loop for BTO380, we need to estimate the internal electric field of the NCs ($E_p$). The particle polarization $P_p(E)$ can be expressed as:

$$P_p(E) = \eta P_{p,NL}(E) + \varepsilon_0(\varepsilon_{p,L} - \varepsilon_{PVP})E_{in} \tag{3}$$

The actual internal field $E_p$ is the sum of the Onsager field $E_{in}$ and the depolarization field from the dipole orientation generated by the single particle polarization $P_{\text{p,NL}}(E)$. The Onsager local field inside the BTO380 particle is:

$$E_{\text{in}} = \frac{3E}{\varepsilon_{\text{p,L}} + 2\varepsilon_{\text{PVP}}} \quad (4)$$

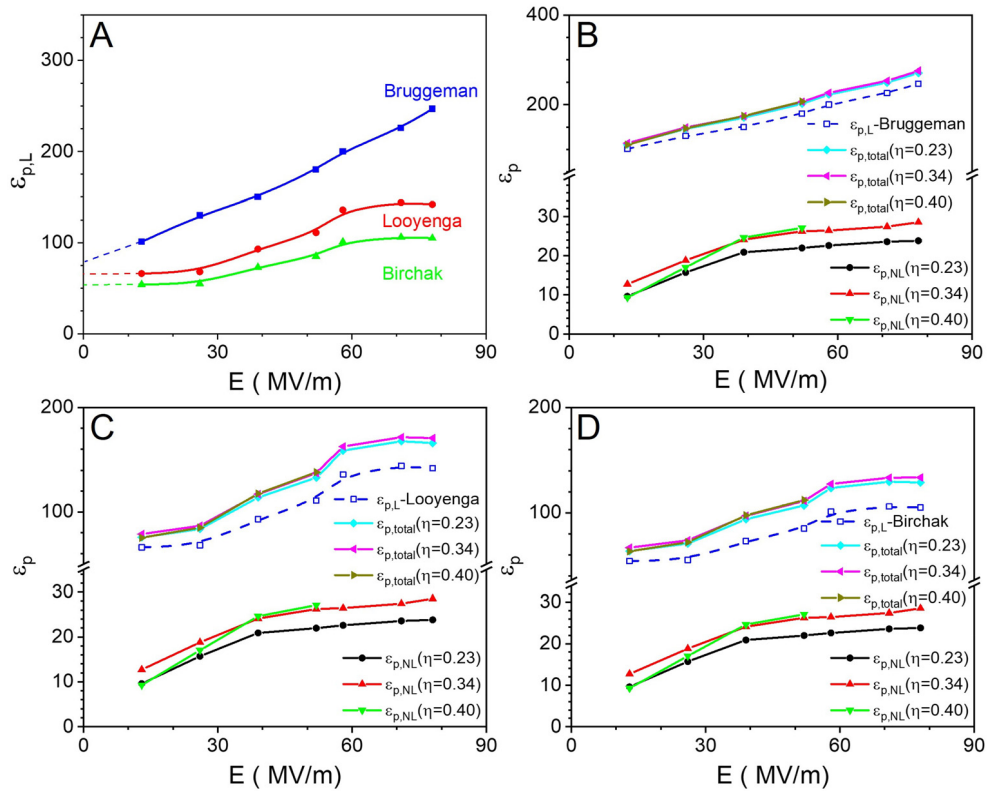

**Fig. 6** (A) Calculated linear dielectric constant $\varepsilon_{p,L}$ of BTO380 NCs using various mixing rules. Nonlinear dielectric constant $\varepsilon_{p,NL}$, linear dielectric constant $\varepsilon_{p,L}$, and total dielectric constant $\varepsilon_p$ of BTO380 NCs calculated using the (B) Bruggeman, (C) Looyenga, and (D) Birchak mixing rules.

$E_{\text{p}}$ is the actual internal field inside the BTO NCs, which is defined as:

$$E_{\text{p}} = E_{\text{in}} - \frac{1}{3\varepsilon_0}\frac{P_{\text{p,NL}}(E)}{\eta} \quad (5)$$

where $P_{\text{p,NL}}(E)$ can be written as

$$\frac{P_{\text{p,NL}}(E)}{\eta} = \varepsilon_0(\varepsilon_{\text{p,NL}} - 1)E_p \quad (6)$$

so, the particle internal field $E_{\text{p}}$ can be rewritten as:

$$E_{\text{p}} = E_{\text{in}} - \frac{\varepsilon_{\text{p,NL}}(E) - 1}{3}E_{\text{p}} \quad (7)$$

or, the Onsager local field becomes:

$$E_{\text{in}} = E_p\frac{\varepsilon_{\text{p,NL}}(E) + 2}{3} \quad (8)$$

Finally, the actual internal field $E_{\text{p}}$ considers the induced polarization of the NCs through the Onsager local field model and also accounts for the orientational depolarization field of the internal dipoles inside the NCs,

$$E_p = \frac{3}{\varepsilon_{\text{p,L}} + 2\varepsilon_{\text{PVP}}}\frac{3}{\varepsilon_{\text{p,NL}}(E) + 2}E \quad (9)$$

Using these equations, the $P_{\text{p}}$–$E_{\text{p}}$ loops of individual BTO380 NCs are plotted in Fig. 7, using the BTO380 permittivities calculated with different mixing rules. Fig. 7A–C represents the $P_{\text{p}}$–$E_{\text{p}}$ loops obtained for the BTO380 NCs using the Bruggeman mixing rule for PVP/BTO380-0.23, PVP/BTO380-0.34, PVP/BTO380-0.40, respectively. As the poling electric field increased, the coercive field $E_{\text{c}}$ approached 0.35 MV m$^{-1}$ and the remanent polarization $P_{\text{r}}$ reached 12.3 mC m$^{-2}$. Fig. 7D–F show the $P_{\text{p}}$–$E_{\text{p}}$ loops extracted from the Looyenga mixing rules with the $E_{\text{c}}$ increased to 0.51 MV m$^{-1}$.

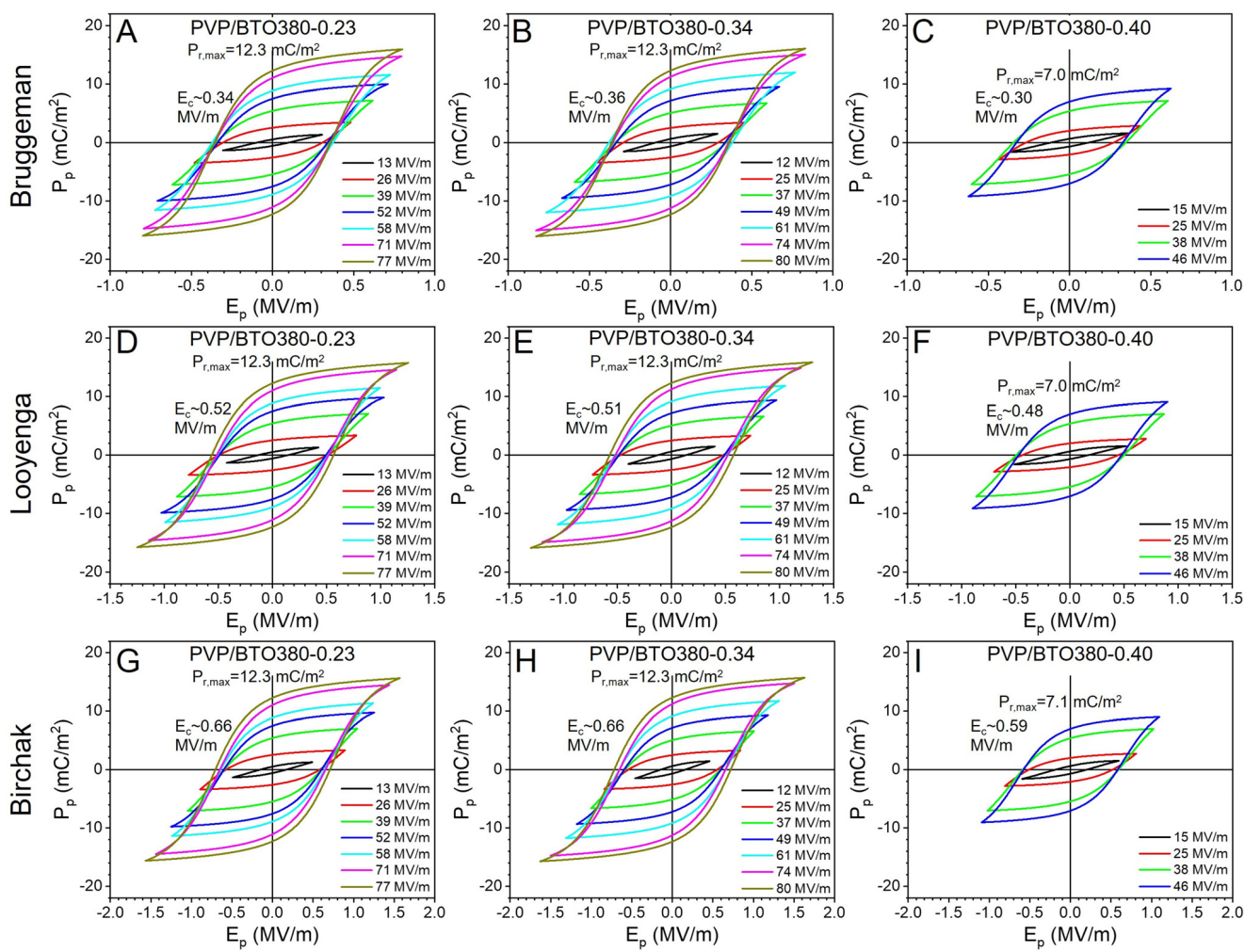

**Fig. 7** Bipolar $P_{\text{p}}$–$E_{\text{p}}$ loops of BTO380 NCs using the (A–C) Bruggeman, (D–F) Looyenga, and (G–I) Birchak mixing models from the PVP/BTO380 composites with different $\eta$ values: (A, D and G) 0.23, (B, E and H) 0.34, and (C, F and I) 0.40.

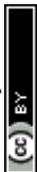

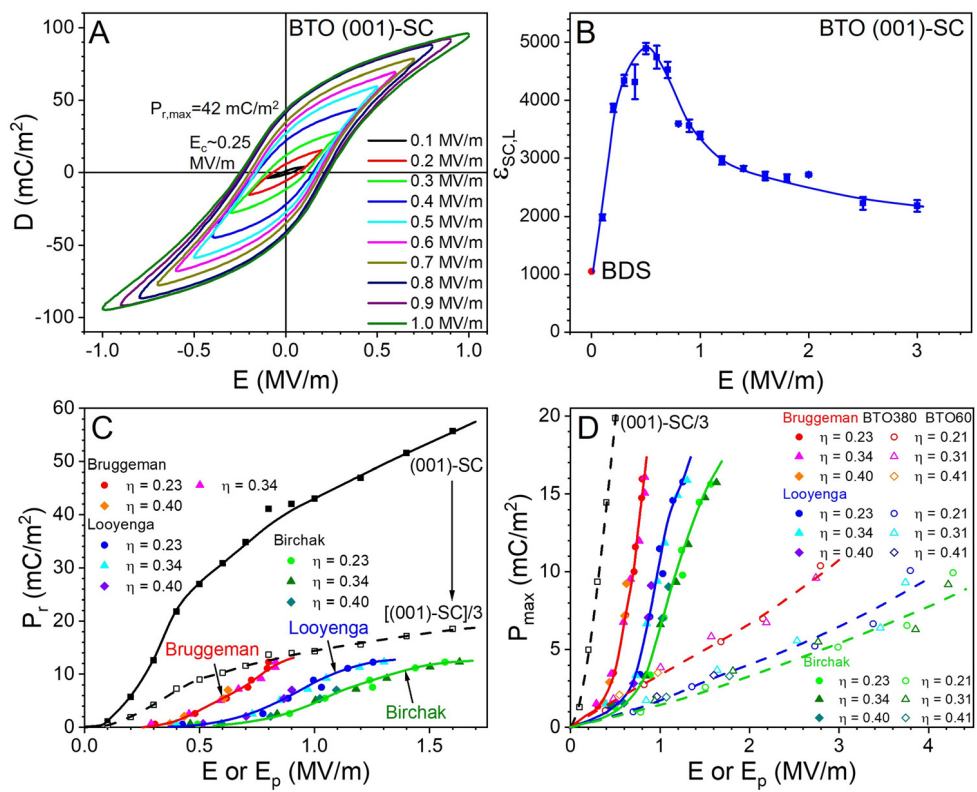

**Fig. 8** (A) Bipolar *D–E* loops of the BaTiO$_3$ (001)-SC at room temperature. (B) Linear permittivity ($\varepsilon_{SC,L}$) as a function of the applied electric field. The poling frequency is 10 Hz with a sinusoidal waveform. (C) The remanent polarization ($P_r$) as a function of the poling field *E* for the (001)-SC or the internal field $E_p$ in the BTO380 NCs calculated from different mixing rules. (D) The maximum polarization ($P_{max}$) as a function of *E* for the (001)-SC or $E_p$ in BTO380 and BTO60 NCs calculated from different mixing rules.

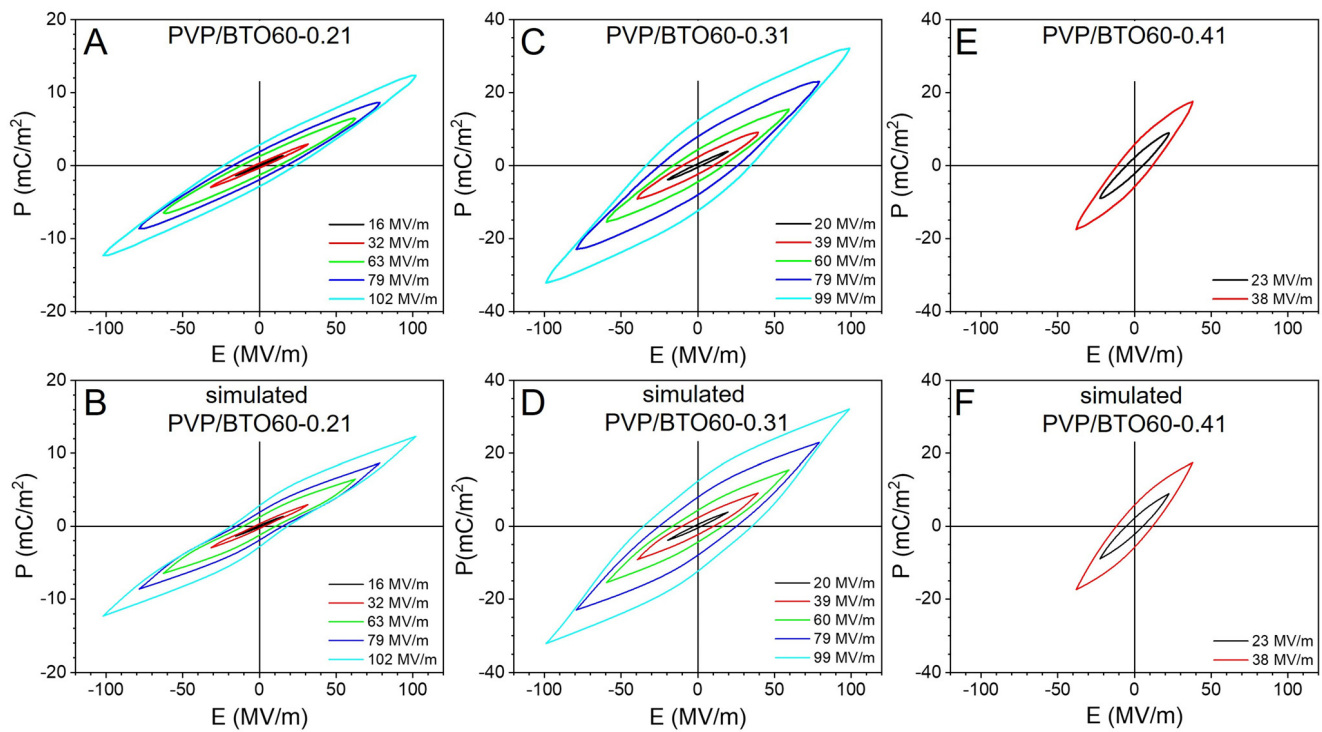

**Fig. 9** Bipolar *P–E* loops of the PVP/BTO60 composites with different BTO packing fractions at room temperature: (A) PVP/BTO60-0.21, (C) PVP/BTO60-0.31, and (E) PVP/BTO60-0.41. The corresponding simulated *P–E* loops are shown in (B), (D), and (F), respectively. The poling frequency is 10 Hz with a sinusoidal waveform.

Fig. 7G–I show the $P_p$–$E_p$ loop extracted from the Birchak mixing rule and the $E_c$ increased to 0.66 MV m$^{-1}$. The increased $E_p$ and $E_c$ for the Looyenga and Birchak loops could be attributed to the lower linear dielectric constants extracted for the BTO380 NCs. This is the first time that the $P$–$E$ loops for ferroelectric BTO NCs were extracted from the $P$–$E$ loops of their nanocomposites.

### Comparison with a multi-domain BTO (001)-SC

As reported previously, the 950 °C-annealed BTO380 NCs were single crystals.[15] For comparison, bipolar electric displacement–electric field ($D$–$E$) loops for a tetragonal BTO (001)-SC were obtained at room temperature and 10 Hz. As shown in Fig. 8A, the values of spontaneous polarization $P_s$, $P_r$, and $E_c$ increased with increasing poling electric field. At 1.0 MV m$^{-1}$, $E_c = 0.25$ MV m$^{-1}$. The $D$–$E$ loops under higher applied fields ($E > 1.0$ MV m$^{-1}$) are plotted in Fig. S2.† In this figure, the $P_s$ value of the BTO (001)-SC was significantly lower than the theoretical $P_s$ value of 260 mC m$^{-2}$,[31] and this could be attributed to the multidomain structure of the (001)-SC.

The linear permittivity ($\varepsilon_{SC,L}$) of the BTO (001)-SC was calculated from the deformational polarization during the initial reverse poling process following previous reports.[24] The results are shown in Fig. 8B. The $\varepsilon_{SC,L}$ peaked around 0.5 MV m$^{-1}$ with a value of 4890. Afterwards, the $\varepsilon_{SC,L}$ gradually decreased to 2500 at 3.0 MV m$^{-1}$. The peak $\varepsilon_{SC,L}$ of 4890 was significantly higher than the value (1050) obtained by broadband dielectric spectroscopy at room temperature and 10 Hz (see Fig. S2B†).[15]

The changes of $P_r$ with the applied electric field $E$ for BTO (001)-SC or the calculated $E_p$ in BTO380 NCs are shown in Fig. 8C. For the BTO (001)-SC, a rapid increase of $P_r$ was observed before 0.5 MV m$^{-1}$, which corresponded to the maximum $\varepsilon_{SC,L}$ in Fig. 8B. With further increase of the poling field, the $P_r$ monotonically increased. We consider that the original (001)-SC was multi-domained with 180° domains. Upon electric poling, these multi-domains quickly switched from the antiparallel orientation to the parallel orientation, providing the maximum $\varepsilon_{SC,L}$. With further increase of the poling field, the parallelly oriented domains gradually grew and merged into larger domains, leading to further increased $P_r$ and $E_c$ (see Fig. S2†). Compared with the (001)-SC, the $P_r$ values for the BTO380 MCs appeared to be much smaller, regardless of which mixing rule was used to estimate the internal $E_p$. We consider that the lower $P_r$ should largely originate from the random orientation of the BTO380 NCs in the composites. Considering the probability of $c$-axis orientation along the poling field direction being 1/3, the $P_r$/3 curve of the (001)-SC is plotted in Fig. 8C. As we can see, this curve matched better with the maximum $P_r$ of 12.3 mC m$^{-2}$ for the BTO380 NCs. Another reason for the lower $P_r$ could be attribu-ted to the early breakdown of the composite films. If the composite films did not breakdown early and a higher field could be reached, the $P_r$ of BTO380 NCs could be even higher. Meanwhile, the $E_p$ values calculated from the Bruggeman equation were closer to the poling electric field for the $P_r/3$ curve of the (001)-SC. Fig. 8D shows the $P_{max}$ as a function of the applied field E or the internal field in the BTO380 NCs, $E_p$. Again, the Bruggeman curve was the closest to the (001)-SC/3 curve. Therefore, we consider that the $P_p$–$E_p$ loops based on the Bruggeman mixing rule were more realistic for the BTO380 NCs (see Fig. 7A–C).

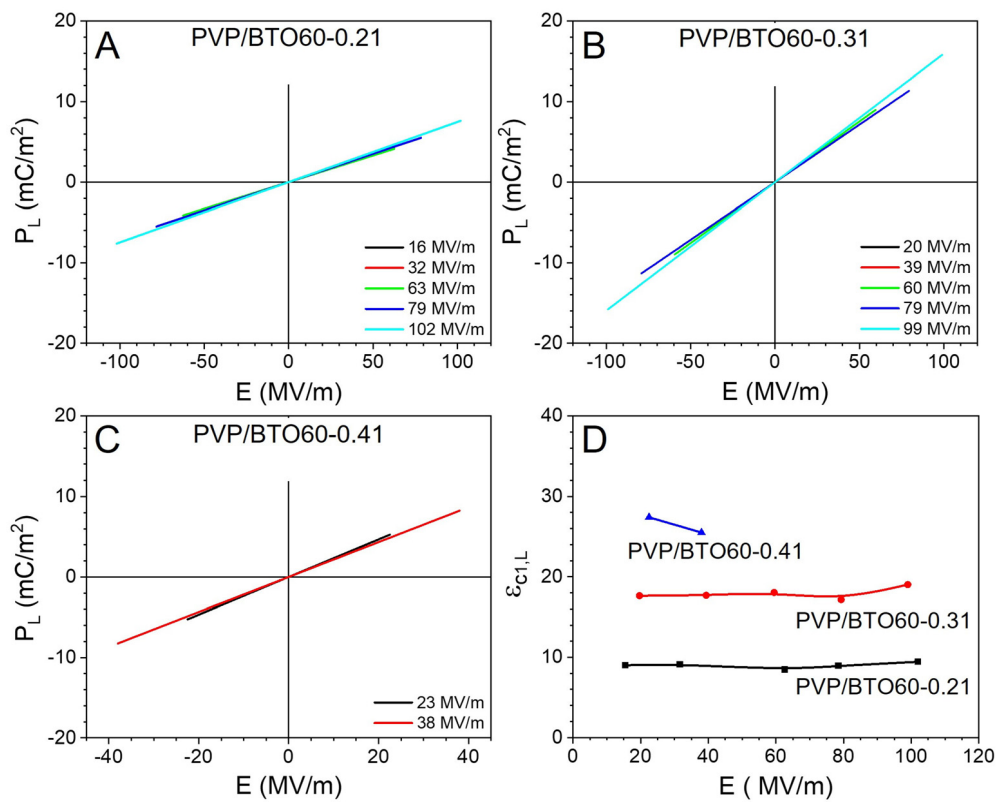

**Fig. 10** Deconvoluted linear $P_L$–$E$ loops of various PVP/BTO60 nanocomposites: (A) PVP/BTO60-0.21, (B) PVP/BTO60-0.31, and (C) PVP/BTO60-0.41 under different electric fields. (D) Linear dielectric constants of various PVP/BTO60 nanocomposites calculated from the corresponding linear $P_L$–$E$ loops.

### Deconvolution of experimental P–E loops for PVP/BTO60 nanocomposites

It is not surprising that the PVP/BTO380 nanocomposite films have dielectric nonlinearity because the BTO380 NCs are ferroelectric in nature. What will happen if the ferroelectric BTO380 NCs are replaced with the paraelectric BTO60 NCs for the PVP nanocomposite? Experimental P–E loops are shown in Fig. 9A, C and E for the PVP/BTO60 nanocomposites with $\eta$ = 0.21, 0.31, and 0.41, respectively. Supposedly, as BTO60 NCs were paraelectric and PVP was a linear dielectric, slim P–E loops should be observed. Nevertheless, broad P–E loops were observed for the PVP/BTO60 nanocomposite films, especially for high packing fractions, i.e., $\eta$ > 0.3. The same theoretical deconvolution procedure was carried out and the simulated P–E loops for the PVP/BTO60 nanocomposites are plotted in Fig. 9B, D and F, respectively. The simulated loops fitted well with the experimental loops.

Again, the experimental P–E loops of the PVP/BTO60 nanocomposites were deconvoluted into three components, namely, $P_L$, $P_p$, and $P_{int}$. The deconvoluted linear $P_L$–E loops are shown in Fig. 10A–C for the PVP/BTO60 nanocomposites with $\eta$ = 0.21, 0.31 and 0.41 under different poling electric fields, respectively. From the slopes of the linear $P_L$–$E$ loops, the linear dielectric constants $\varepsilon_{c1,L}$ of the PVP/BTO60 nanocomposites were calculated and are plotted in Fig. 10D. The $\varepsilon_{c1,L}$ values remained nearly constant under different electric fields for PVP/BTO60-0.21 and PVP/BTO60-0.31 up to 100 MV/m: $\varepsilon_{c1,L} = 18$ for PVP/BTO60-0.31 and $\varepsilon_{c1,L} = 9$ for PVP/BTO60-0.21. This observation is somewhat different from that of the PVP/BTO380 composites (Fig. 3D), revealing that the dielectric properties of the nanocomposite films could be affected by different crystal structures of BTO fillers (see the results later).

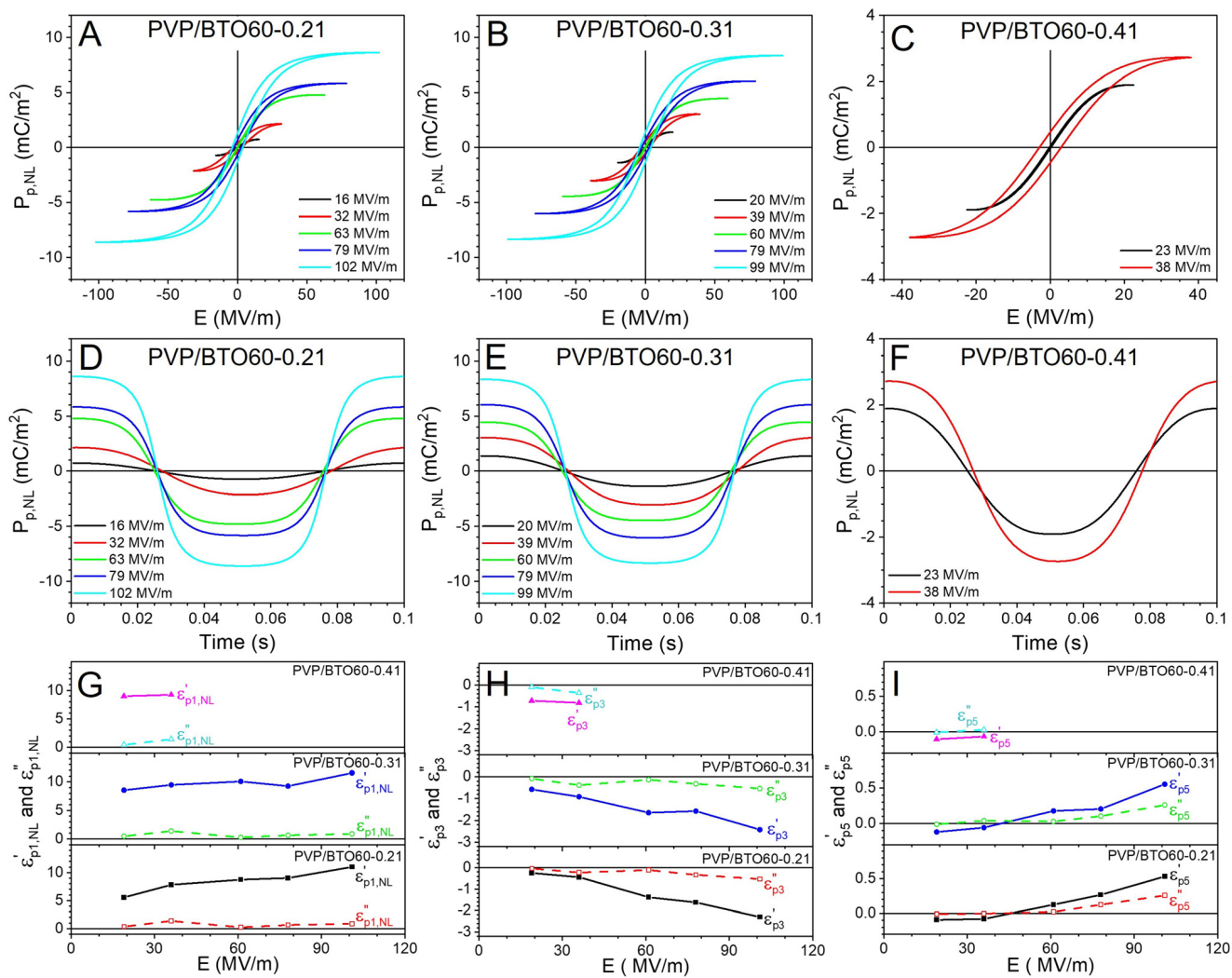

**Fig. 11** (A–C) Bipolar $P_p$–E loops of BTO60 NCs extracted from the P–E loops of the PVP/BTO60 nanocomposites with different BTO packing fractions. (D–F) Nonlinear $P_p(t)$ waves as a function of time for neat BTO60 NCs. After Fourier transform, the nonlinear harmonics, (G) $\varepsilon'_{p1,NL}$ and $\varepsilon''_{p1,NL}$, (H) $\varepsilon'_{p3}$ and $\varepsilon''_{p3}$, and (I) $\varepsilon'_{p5}$ and $\varepsilon''_{p5}$, were calculated from the nonlinear $P_p(t)$ waves.

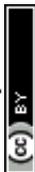

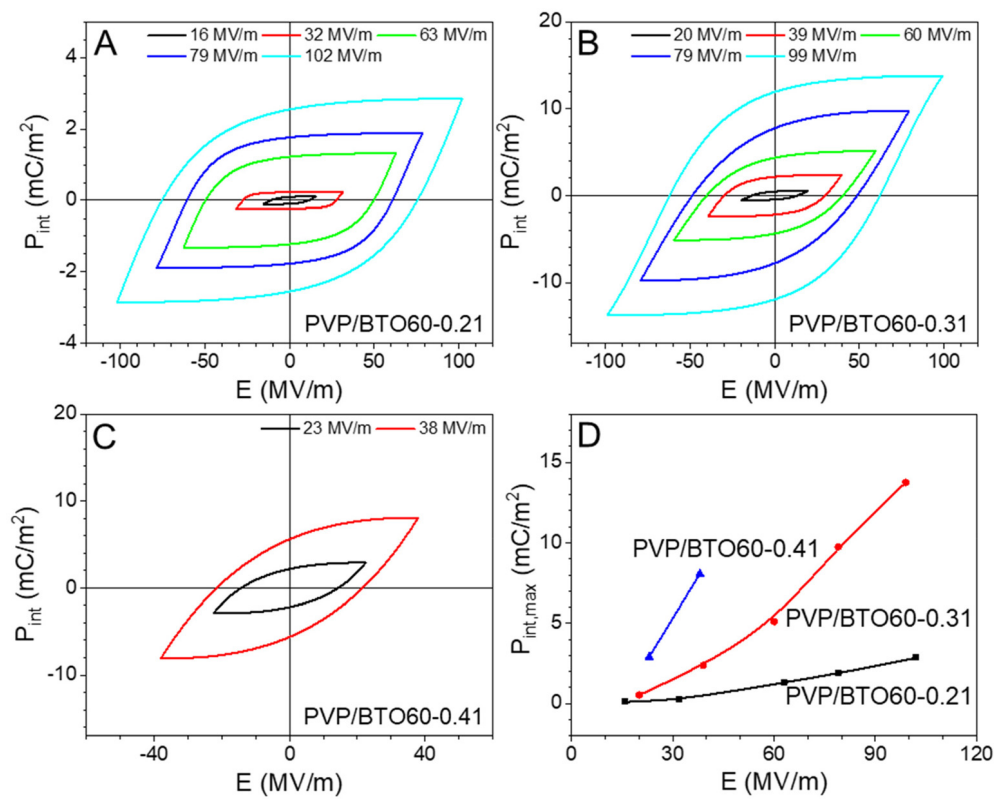

Fig. 12 (A–C) Bipolar particle–particle interaction $P_{int}$–$E$ loops extracted from the experimental $P$–$E$ loops of the PVP/BTO60 nanocomposites with different BTO packing fractions. (D) The corresponding maximum polarization ($P_{int,max}$) under different electric fields.

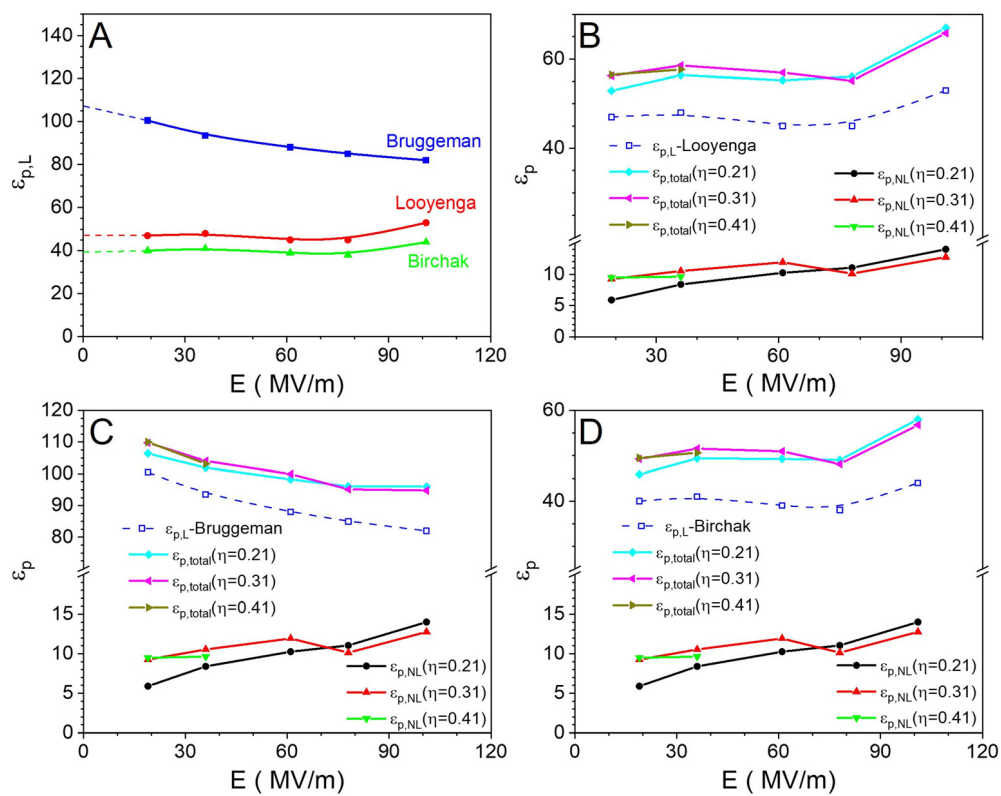

Fig. 13 (A) Calculated linear dielectric constant $\varepsilon_{p,L}$ of BTO60 NCs using various mixing rules. Nonlinear dielectric constant $\varepsilon_{p,NL}$, linear dielectric constant $\varepsilon_{p,L}$, and total dielectric constant $\varepsilon_p$ of BTO60 NCs calculated using the (B) Bruggeman, (C) Looyenga, and (D) Birchak mixing rules.

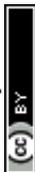

Compared to the broad ferroelectric loops of the BTO380 NCs, the $P_p$–$E$ loops of the BTO60 NCs appeared to be slim with an $S$ shape, which is typical for paraelectric BTO (see Fig. 11A–C). The saturation polarization $P_{max}$ increased with increasing applied electric field; however, the $P_{max}$ of BTO60 was much smaller than that of BTO380 at the same electric field. For example, the $P_{max}$ was ~8.6 mC m$^{-2}$ for PVP/BTO60-0.21 at 79 MV m$^{-1}$, whereas the $P_{max}$ was ~15 mC m$^{-2}$ for PVP/

BTO380-0.23 at 77 MV m$^{-1}$. The same procedure of nonlinear dielectric analysis was applied for the PVP/BTO60 nanocomposites. The $P_p(t)$ waves under different applied fields are plotted in Fig. 11D–F for $\eta$ values of 0.21, 0.31, and 0.41, respectively. After Fourier transform, the nonlinear $\varepsilon'_{pn}$ and $\varepsilon''_{pn}$ ($n = 1, 3, 5$) for BTO60 were calculated and the results are shown in Fig. 11G–I. The values of $\varepsilon''_{p1,NL}$ and $\varepsilon''_{p1,NL}$ of BTO60 were smaller compared to those of BTO380 due to their paraelectricity.

Although the paraelectric $P_p$–$E$ loops of the BTO60 NCs were dramatically different from the ferroelectric $P_p$–$E$ loops of BTO380 NCs, the $P_{int}$–$E$ loops appeared to be quite similar; see Fig. 12A–C. Specifically, the $P_{int,max}$ increased with the applied electric field. This result is surprising because the paraelectric BTO60 NCs could form ferroelectric "clusters" and induce significant nonlinearity under high poling fields. The $P_{int,max}$ values are shown in Fig. 12D, where the $P_{int,max}$ increased with the applied electric field for all the PVP/BTO60 nano-composites. However, the $P_{int,max}$ values of BTO60 were lower than those of BTO380 (comparing Fig. 5 and 12), indicating a weaker particle–particle interaction. Meanwhile, at the same high electric field (>40 MV m$^{-1}$), the $P_{int,max}$ was significantly higher than the $P_p$ value of the BTO60 NCs, suggesting that the field-induced transient ferroelectricity from the particle–particle interactions among the BTO60 NCs was the major contributor to nonlinearity in the linear dielectric polymer/paraelectric ceramic nanofiller composites.

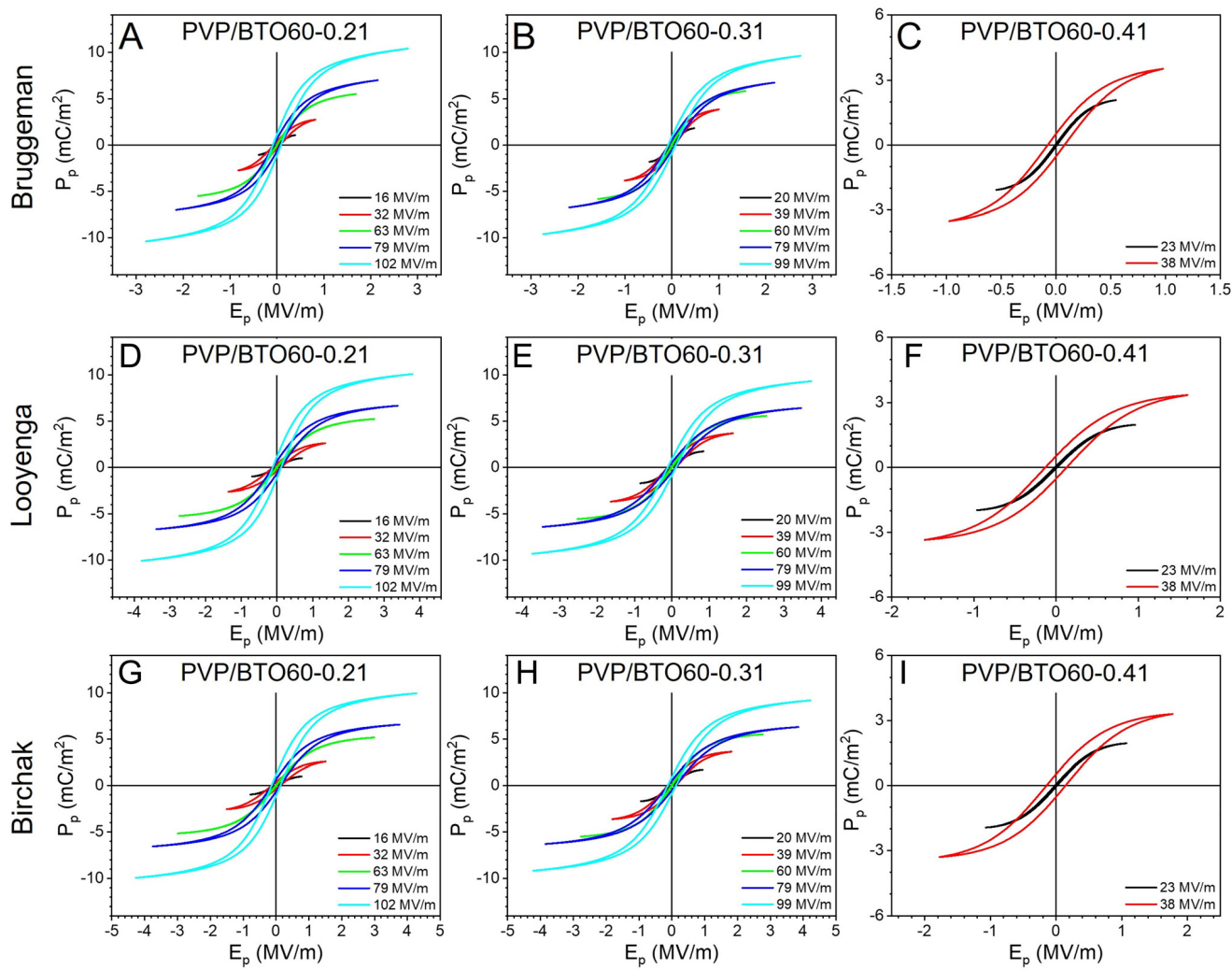

**Fig. 14** Bipolar $P_p$–$E_p$ loops of BTO60 NCs using the (A–C) Bruggeman, (D–F) Looyenga, and (G–I) Birchak mixing models for the PVP/BTO60 nanocomposites with different $\eta$ values: (A, D and G) 0.21, (B, E and H) 0.31, and (C, F and I) 0.41.

Using different mixing rules, the linear $\varepsilon_{p,L}$ of the BTO60 NCs can be extracted from the linear dielectric constants of the PVP/BTO60 nanocomposites (see Fig. 13A). The linear $\varepsilon_{p,L}$ of the BTO60 NCs slightly decreased from 100 to 80 with increasing the poling electric field from 20 to 100 MV m$^{-1}$ for the Bruggeman mixing rule. However, the linear $\varepsilon_{p,L}$ remained nearly constant at 46 for the Looyenga mixing rule and 40 for the Birchak mixing rule. The $\varepsilon_{p,L}$ values obtained by the Bruggeman mixing rule were similar to what we reported for the BTO60 NCs.[15] By adding the nonlinear dielectric constant, $\varepsilon_{p,NL}$, the total dielectric constants calculated using the Bruggenman, Looyenga, and Birchak mixing rules were also obtained, see Fig. 13B–D, respectively. Apparently, these dielectric constant values were significantly lower than those of the BTO380 NCs, as seen in Fig. 6.

Using eqn (3)–(9), the $E_p$ value was calculated based on the total dielectric constant of the BTO60 NCs. The $P_p$–$E_p$ loops of the paraelectric BTO60 NCs calculated using the Bruggeman, Looyenga, and Birchak mixing rules are plotted in Fig. 14 for various PVP/BTO60 nanocomposites. Again, slim paraelectric $P_p$–$E_p$ loops were observed. The maximum polarization $P_{max}$ values *versus* $E_p$ for different mixing rules are summarized in Fig. 8D. As we can see, the $P_{max}$–$E_p$ curves for different nanocomposites collapsed onto the same line, indicating that the $P_p$–$E_p$ loops were independent of the PVP/BTO60 nanocomposite compositions. Meanwhile, the $P_{max}$–$E_p$ curve from the Bruggeman mixing rule was the closest to the (001)-SC/3 curve. Again, we therefore consider that the $P_p$–$E_p$ loops from the Bruggeman mixing rule might be better to represent the BTO60 NCs.

## Conclusions

In summary, by capitalizing on the deconvolution of experimental $P$–$E$ loops, the $P_p$–$E_p$ hysteresis loops of ferroelectric BTO380 and paraelectric BTO60 NCs were determined from their PVP nanocomposites at different packing fractions. Square-shaped ferroelectric loops were obtained for the tetragonal BTO380 NCs. In contrast, slim paraelectric loops were obtained for the cubic BTO60 NCs. Given the difficulty of using PFM in accurately measuring the $P_p$–$E_p$ loops for ferroelectric NCs, the nanocomposite approach was relatively simple and straightforward. Yet, some aspects are worth noting. First, the BTO NCs were randomly oriented in the nanocomposite films, and the breakdown strengths were generally low for nanocomposites with large dielectric constant contrast.[6,29] Therefore, the achievable $P_{p,max}$ appeared to be significantly lower than that of the BTO (001)-SC. Second, various mixing rules have to be used to estimate the internal electric field of BTO NCs. As such, uncertainty could be introduced in the final $P_p$–$E_p$ loops, although the Bruggeman mixing rule seems to be more appropriate than other mixing rules.

In addition, there existed strong particle–particle dipolar interactions, leading to additional ferroelectricity in the experimental $P$–$E$ loops of the PVP/BTO NC composites. This is somewhat comprehensible for the PVP/BTO380 composites given the ferroelectric nature of BTO380 NCs. However, it was unexpected to observe substantial ferroelectric hysteresis in the case of linear PVP/paraelectric BTO60 nanocomposites. We consider that the local flocculation of BTO60 NCs in the nanocomposites was responsible for such strong particle–particle interactions. This result signifies that when aiming for capacitive energy storage applications,[6,32–38] it is advisable to avoid high filler contents in polymer nanocomposites. Alternatively, ultrafine hairy BTO NCs with high dielectric constants[39–41] are suggested to be employed for polymer nanodielectrics at a relatively low filler content. We expect that the densely grafted polymer brushes on these BTO NCs will enable uniform dispersion in polymer matrices to avoid local flocculation of nanoparticles.

## Conflicts of interest

The authors declare no conflict of interest.

## Acknowledgements

L. Z. and Z. L. acknowledge the financial support from the National Science Foundation, Division of Materials Research (DMR), Solid State and Materials Chemistry Program (DMR-1709420). E. A. is thankful for the financial support from the Ministry of Science and Higher Education of the Russian Federation (State Assignment No. 075-01056-22-00).